\documentclass[twocolumn,showpacs,preprintnumbers,amsmath,amssymb]{revtex4}

\usepackage{latexsym}

\usepackage{graphicx}
\usepackage{amssymb}
\usepackage{epstopdf}
\begin{document}

\title{On capillary-gravity waves generated by a slow moving object
}
\author{A. D. Chepelianskii$^{(a,b)}$, F. Chevy$^{(c)}$ and E. Rapha\"el$^{(a)}$}
\affiliation{$^{(a)}$ Laboratoire Physico-Chimie Th\'eorique, UMR CNRS Gulliver 7083, ESPCI, 10 rue Vauquelin, 75005 Paris, France}
\affiliation{$^{(b)}$ Laboratoire de Physique des Solides, UMR CNRS 8502, B\^at. 510, Universit\'e Paris-Sud, 91405 Orsay, France }
\affiliation{$^{(c)}$ Laboratoire Kastler Brossel, ENS, Universit\'e Paris 6, CNRS, 24 rue Lhomond, 75005 Paris, France}

\date{\today}

\begin{abstract}

We investigate theoretically and experimentally the
capillary-gravity waves created by a small object moving steadily
at the water-air interface along a circular trajectory. It is well
established that, for straight uniform motion, no steady waves
appear at velocities below the minimum phase velocity
$c_{\rm{min}} = 23 \; \rm{cm \cdot s}^{-1}$. We show theoretically
that no such velocity threshold exists for a steady circular
motion, for which, even for small velocities, a finite wave drag
is experienced by the object. This wave drag originates from the
emission of a spiral-like wave pattern. Our results are in good
agreement with direct experimental observations of the wave
pattern created by a circularly moving needle in contact with
water. Our study leads to new insights into the problem of animal
locomotion at the water-air interface.
\end{abstract}

\maketitle

Capillary-gravity waves propagating at the free surface of a
liquid are driven by a balance between the liquid inertia and its
tendency, under the action of gravity and surface tension forces,
to return to a state of stable equilibrium \cite{LandauLifshitz}.
For an inviscid liquid of infinite depth, the dispersion relation
relating the angular frequency $\omega$ to the wave number $k$ is
given by ${\omega}^2 = g k + \gamma k^3/\rho$, where $\rho$ is the
liquid density, $\gamma$ the liquid-air surface tension, and $g$
the acceleration due to gravity \cite{Acheson}. The above equation
may also be written as a dependence of the wave velocity $c(k) =
\omega(k)/k$ on wave number: $c(k) = {\left(g/k + \gamma
k/\rho\right)}^{1/2}$.
The dispersive nature of capillary-gravity waves is
responsible for the complicated wave pattern generated at the free
surface of a still liquid by
a moving disturbance such as a partially  immersed
object   ({\it{e.g.}} a boat or an insect) or
an external surface pressure source \cite{Acheson,
Lighthill, Lamb, Rayleigh, Kelvin}.
Since the disturbance expends a power to generate these waves,
it will experience a drag, ${R}_w$, called the
{\it wave resistance} \cite{Lighthill}. In the case of boats and
large ships, this drag is known to be a major source of resistance
and important efforts have been devoted to the design of
hulls minimizing it \cite{Milgram}. The case of objects small relative
to the capillary length $\kappa^{-1} = {\left(\gamma/(\rho g)\right)}^{1/2}$
has only recently been considered
\cite{PGG,Richard,Sun,Chevy}.

In the case of a disturbance moving at {\it constant} velocity
$\boldsymbol V$, the wave resistance $R_w$ cancels out for $V <
c_{\rm{min}}$ where $V$ stands for the magnitude of the velocity,
and $c_{\rm{min}} = (4 g \gamma  /\rho)^{1/4}$ is the minimum of
the wave velocity $c(k)$ given above for capillarity gravity waves
\cite{Lighthill,Lamb,PGG}. For water with $\gamma = 73 \; \rm{mN
\cdot  m^{-1}}$ and $\rho = 10^3 \; \rm{kg \cdot m^{-3}}$, one has
$c_{\rm{min}} = 0.23 \; \rm{m \cdot s^{-1}}$ (room temperature).
This striking behavior of $R_w$ around $c_{\rm{min}}$ is similar
to the well-known Cerenkov radiation emitted by a charged particle
\cite{Cherenkov}, and has been recently studied experimentally
\cite{Browaeys,Burghelea1}. In this letter, we demonstrate that
just like {\it accelerated} charged particles radiate
electromagnetic waves even while moving slower than the speed of
light \cite{Jackson}, an accelerated disturbance experiences a
non-zero wave resistance $R_w$ even when propagating below
$c_{\rm{min}}$. We consider the special case of a uniform circular
trajectory, a situation of particular importance for the study of
whirligig beetles ({\it Gyrinidae}, \cite{Nachtigall1965})
 whose characteristic circular
motion  might facilitate the emission of surface waves that they
are thought to be used for echolocation \cite{Tucker1,Denis1}.
This work is therefore restricted to the effect of a wake
stationary in the rotating frame, and do not consider time
dependent contributions, like vortex shedding \cite{Buhler,Bush}.

We consider the case of an incompressible infinitely deep liquid
whose free surface is unlimited. In the absence of external
perturbation, the free surface is flat and each of its points can
be described by a radius vector $\boldsymbol{r} = (x,y)$ in the
horizontal plane. The motion of a small object along the free
surface disturbs the equilibrium position of the fluid, and each
point of the free surface acquires a finite vertical displacement
$\zeta(\boldsymbol{r})$. Rather than solving the complex
hydrodynamic problem of finding the flow around a moving object,
we consider the displacement of an external pressure source
$P_{ext}(\boldsymbol{r},t)$ \cite{Rayleigh,Kelvin}. The equations
of motion can then be linearized in the limit of small wave
amplitudes \cite{Dias}.

In the frame of this linear-response theory, it is convenient to
introduce the Fourier transforms of the pressure source $\hat
P_{ext}(\boldsymbol{k},t)$ and of the vertical displacement $\hat
\zeta(\boldsymbol{k},t)$ \cite{FourierDefinition}. It can be shown
that, in the limit of small kinematic viscosity $\nu$, the
relation between $\hat \zeta(\boldsymbol{k},t)$ and $\hat
P_{ext}(\boldsymbol{k},t)$ is given by \cite{Richard}

\begin{equation}
\frac{ \partial^2  {\hat \zeta}  }{ \partial t^2 } \; + \; 4 \, \nu \, k^2Ê\,  \frac{ \partial {\hat \zeta} }{ \partial t } \; + \; \omega^2(k) \, {\hat \zeta} \;
= \;  - \frac{ k {\hat P}_{ext}(\boldsymbol{k},t)}{ \rho}
\label{eqzeta}
\end{equation}

In this letter we assume that the pressure source has radial
symmetry and that the trajectory $\boldsymbol{r}_0(t)$ of the
object is circular, namely : $\boldsymbol{r}_0(t) = \mathcal{R} \,
(\cos(\Omega t), \sin(\Omega t))$. Here $\mathcal{R}$ is the
circle radius, and $\Omega$ is the angular frequency. The linear
velocity of the object is then given by $V = \mathcal{R} \,
\Omega$. With these assumptions, the external pressure field is
$P_{ext}(\boldsymbol{r}, t) = P_{ext}(|\boldsymbol{r}
-\boldsymbol{r}_0(t)|$, yielding in Fourier space $\hat
P_{ext}(\boldsymbol{k}, t) = \hat P_{ext}(k) e^{- i
\boldsymbol{k}.\boldsymbol{r}_0(t)} $. Since the right hand side
of  Eq.~(\ref{eqzeta}) is periodic with frequency $\Omega$, it is
possible to find its steady state solution by expanding the right
hand side into Fourier series. The problem then becomes equivalent
to the response of a damped oscillator to a sum of periodic forces
with frequencies $n \Omega$, where $n$ is an integer. The vertical
deformation at any time $t$ can then be reconstructed by
evaluating the inverse Fourier transform. For the particular case
of uniform circular motion, the time dependence is rather simple.
Indeed, in steady state, the deformation profile rotates with the
same frequency $\Omega$ as the disturbance. Therefore, in the
rotating frame, $\zeta$ depends on the position $\boldsymbol{r}$
only. The analytical expression of $\zeta(\boldsymbol{r})$ in
cylindrical coordinates $(x, y) = r (\cos \phi, \sin \phi)$ is
given by

\begin{align}
\zeta(r, \phi) &= \sum_{n = -\infty}^{\infty}   e^{i n \phi}  \int \frac{k^2 d k}{ 2 \pi \rho} \frac{{\hat P}_{ext}(k) J_n(k r) J_n(k \mathcal{R})}{n^2 \Omega^2 - \omega^2(k) + 4 i n \nu k^2 \Omega }
\label{zetaexpr}
\end{align}

\noindent where $J_n$ is $n$-th order Bessel function of the first
kind. The summation index $n$ is directly related to the $n$-th
Fourier harmonic of the periodic function $ e^{- i
\boldsymbol{k}.\boldsymbol{r}_0(t)} $ and, since the problem is
linear, the contributions of all the harmonics add together.

The knowledge of the exact structure of the wave pattern is
precious, but a quantitative measurement of the wave resistance is
needed in order to understand, for example, the forces developed
by small animals moving at the surface of water. In the case of
the circular motion under study, the wave resistance $R_{\rm w}$
can be calculated from its average power $P_{\rm w} =  - \int d^2
r \left\langle P_{ext}(\boldsymbol{r}, t) \frac{\partial
\zeta(\boldsymbol{r}, t)}{\partial t}\right\rangle $ by
$R_w=P_{\rm w}/V$. Using the Fourier expansion of $\zeta$, one
then obtains in the limit $\nu \kappa / c_{\rm{min}} \rightarrow
0$ (for water, $\nu \kappa / c_{\rm{min}} \sim 10^{-3}$):

\begin{align}
R_w(V, \mathcal{R})   =\sum_{n > 0} \frac{n}{\rho \mathcal{R}}
\frac{(k_n J_n(k_n \mathcal{R}) {\hat P}_{ext}(k_n))^2}{ \left(
\frac{d \, \omega^2}{d k} \right)_{k_n}} \label{rw}
\end{align}

\noindent where $k_n$ is the unique solution of the equation
$\omega(k_n) = n \, \Omega$ (the notation $R_w(V, \mathcal{R})$
stresses the dependence of $R_w$ on the velocity magnitude and on
the trajectory radius).
 Equation~(\ref{rw}) shows that the
wave resistance $R_w$ takes the form of a sum $R_w = \sum_{n>0}
A_n$, where the $A_n$ are positive numbers that measure the
contribution of each Fourier mode of the external pressure source
(with frequency $n \, \Omega$) to the wave resistance.

\begin{figure}
\begin{center}
\includegraphics[width= 0.8 \columnwidth]{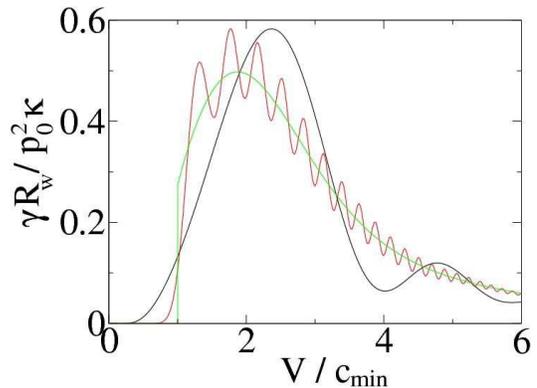}
\end{center}
\caption{(Color online) Plot of the wave resistance $R_w$ in units
of $p_0^2 \kappa/\gamma$, as a function the reduced velocity $V /
c_{\rm{min}} = \mathcal{R} \,\Omega / c_{\rm{min}}$ for different
ratios between the trajectory radius $\mathcal{R}$, and the object
size $b$, as predicted by Eq.~(\ref{rw}). The red curve
(presenting many oscillations) corresponds  to $\mathcal{R} / b =
100$, while the black one (with fewer oscillations) corresponds to
$\mathcal{R} / b = 10$. The green curve displaying a typical
discontinuity at $V = c_{\rm{min}}$ is the wave drag for a
straight uniform motion with velocity $V$  \cite{PGG}. The object
size, $b$, was set to $b = 0.1 \, \kappa^{-1}$. } \label{wavefig1}
\end{figure}

A numerical calculation of the wave resistance is presented in
Fig. \ref{wavefig1} for a pressure source $\hat P_{ext}(k) = p_0
\exp(- k b)$, where $p_0$ is the total force exerted on the
surface and $b$ is the typical object size \cite{PressureField}.
As observed, Eq.~\ref{rw} differs significantly from the original
prediction on the wave drag in the case of a straight uniform
motion with velocity $V$ \cite{PGG,Chevy} given by

\begin{equation}
R_{w,l}(V) = \int_0^{\infty} \frac{k d k}{2 \pi \rho}
\frac{{\hat{P}_{ext}}^2(k) \, \theta(V - c(k))}{ V^2 \sqrt{1 -
(c(k) / V)^2}}, \label{eqnLinear}
\end{equation}

\noindent where $\theta(.)$ is the Heavyside function and $c(k) =
\omega(k) / k$ is the phase velocity. Most notably, the wave drag
for a circular motion is non-zero for all velocities, even for
$V<c_{\rm min}$ where wave-resistance vanishes exactly in the case
of a linear motion and this effect is far from negligible: for
$\mathcal{R} / b = 10$ and at velocities as slow as $V/c_{\rm
min}\sim 0.6$, the wave drag is still one fifth of that applied to
an object moving linearly at $V/c_{\rm min}=1$. The radiation of
waves by an accelerated particle should not be surprising and
actually is a very general phenomenon that can be observed for
instance in electromagnetism ({\em bremsstrahlung}) or in general
relativity (Zeldovich-Starobinsky effect \cite{Zeldovitch71}).
Mathematically, the fact that, for a circular motion, the wave
resistance is finite even below $c_{\rm min}$ can be understood as
follows. In the case of uniform motion, all the wavenumbers such
as $c(k) < V$ contribute to the wave drag, whereas for circular
motion this is the case for only a discrete set of wavenumbers
$k_n$. While the condition $c(k) < V$ can be satisfied only when
$V > c_{\rm{min}}$, the equations for the wavenumber $k_n$,
$\omega(k_n) = n V / \mathcal{R}$, have positive solutions for any
velocity $V$. These wavenumbers $k_n$ create finite contributions
$A_n > 0$ to the wave drag. Therefore for a circular trajectory a
finite wave drag exists at any velocity $V > 0$; for the same
reasons $R_w$ is also continuous at $V = c_{\rm{min}}$. Moreover,
the wave resistance develops a small oscillating component as a
function of the velocity $V$. It originates from the oscillatory
behavior of Bessel functions and will be analyzed more thoroughly
in a future publication. Finally, we note that despite these
striking differences Eqn. (\ref{rw}) and (\ref{eqnLinear}) should
coincide in the limit of a large trajectory radius $\mathcal R$.
We confirmed this behavior by checking both analytically
\cite{Private} and numerically  that in the limit $\mathcal{R}
\rightarrow \infty$, $R_w(V, \mathcal{R}) \rightarrow R_{w,l}(V)$.
However even if the circular wave drag $R_w(V, \mathcal{R})$ is
close to $R_{w,l}(V)$ starting from $\mathcal{R}/b \sim 10$,
important differences remain even up to very large values of
$\mathcal{R}/b$ such as $\mathcal{R}/b \sim 100$.

\begin{figure}
\begin{center}
\includegraphics[width= 0.5
\columnwidth]{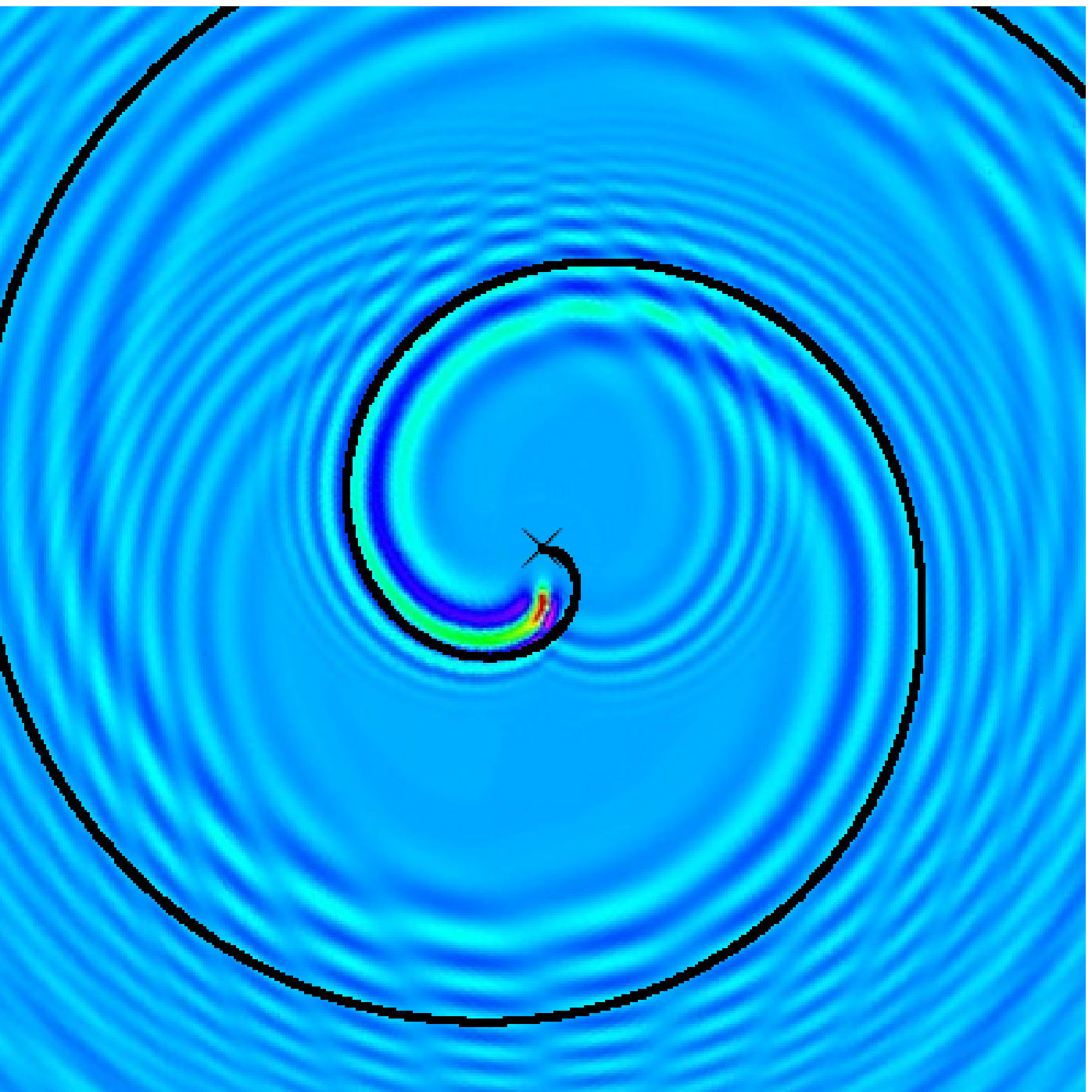}\hfill\includegraphics[width=0.5
\columnwidth]{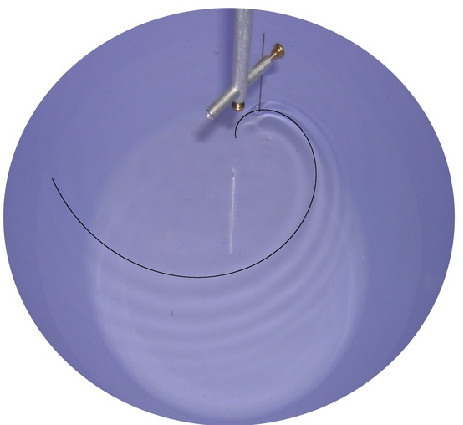}
\end{center}
\label{wavefig3} \caption{(color online) Wave radiation for $V
\approx 21$cm /s $\approx 0.9 c_{\rm{min}}$ with a radius
$\mathcal{R} \approx 2.7$cm $\approx 9 \kappa^{-1}$ Left: Color
diagram of the surface deformation $\zeta(\mathbf{r})$ computed
numerically from Eq.~(\ref{zetaexpr}). This image represents a
square region of size $400 \kappa^{-1}$ around the center of
rotation, red color corresponds to maximal $\zeta(\mathbf{r})$
values, while green corresponds to minimal values of
$\zeta(\mathbf{r})$. The cross indicates the center of the
trajectory and the moving object is located in the region of
highest deformation. Right: Photography of the wave crests
generated on a water surface by a needle rotating at a velocity .
On both pictures, the black curve represents the the Archimedean
spiral of radius given by Eq.~(\ref{rspiral}. }
\end{figure}


Figure~2 represents the wave crest pattern (computed numerically
form Eq.(\ref{zetaexpr})) at the origin of this finite wave drag.
It exhibits characteristic concentric Archimedean spirals (also
known as arithmetic spirals) of the form $r = a \phi + r_0$. This
can be understood from our theoretical results as follows. In a
first estimation, one can assume that the integrals in equation
Eq.~(\ref{zetaexpr}) are dominated by the contribution of the
poles at $k = k_n$. Thus $\zeta(\boldsymbol{r})$ can be written as
$\zeta(\boldsymbol{r}) \sim \frac{1}{\sqrt{r}} \sum_{n} B_n e^{i
(n \phi - k_n r)} $, where we have used the asymptotic development
of $J_n(k_n r)$ at large distances $r$ and $B_n$ are complex
coefficients that do not depend on the position $\boldsymbol{r} =
r(\cos \phi, \sin \phi)$. By separating the contribution of the
different modes in the relation ${\boldsymbol F}(t) = - \int d^2 r
P_{ext}(\boldsymbol{r}, t) \boldsymbol{\nabla}
\zeta(\boldsymbol{r}, t)$, one finds that $B_n$ is proportional to
$A_n$ (where, as defined earlier, the positive coefficients $A_n$
measure the contribution of each Fourier mode to the wave drag:
$R_w = \sum_{n > 0} A_n$). One can show that in the regime of
small object sizes $\kappa b \ll 1$, the proportionality constant
between $B_n$ and $A_n$ depends only weakly on the Fourier mode
number $n$; thus, one has $\zeta(\boldsymbol{r}) \propto
\frac{1}{\sqrt{r}} \sum_{n} A_n e^{i (n \phi - k_n r)}$. We have
checked numerically that in the regime $V < c_{\rm{min}}$, the
distribution of the coefficients $A_n$ is usually peaked around $n
\sim \kappa \mathcal{R}$. For example, for $\kappa \mathcal{R} =
10$ and $\kappa b = 0.1$, $A_n$ is peaked around $n = 10$ for
velocities $V$ in the interval $(c_{\rm{min}}/2, c_{\rm{min}})$.
The wave-crests are given by the lines of constant phase $n \phi -
k_n r = const$ of the dominant mode $n = \kappa \mathcal{R}$,
leading to the following expression for $a$:

\begin{align}
a \approx \frac{\kappa \mathcal{R}}{k(\omega = \kappa V)}
\label{rspiral}
\end{align}
where $k(\omega)$ is the inverse function of $\omega(k)$. An
interesting special case of the formula Eq.~(\ref{rspiral})
corresponds to $V = c_{\rm{min}}$, for which one obtains $a
\approx \mathcal{R}$. The spiral predicted by Eq.~(\ref{rspiral})
is in very good agreement with the exact numerical results
(Eq.~(\ref{zetaexpr})), as can be seen in Fig.~2.

We have also compared our theoretical approach with experimental
results obtained using a one millimeter wide stainless steel
needle immersed in a 38~cm wide water bucket. The needle was
rotated on circular trajectories of various radii and angular
velocities. Since direct measurement of wave drag, and in
particular comparison with theory, is non-trivial even for a
linear motion \cite{Browaeys,Burghelea1}, we restricted ourselves
to the study of the wake itself.
 A typical wave
pattern obtained by this method is shown on Fig.~3 for
$\mathcal{R} \approx 2.7 {\rm cm}$ and $\Omega \approx 2 \pi
\times 1.2 \, {\rm Hz}$ (corresponding to $V/c_{\rm{min}} \approx
0.9$) and unambiguously demonstrates the existence of a wake at
velocities smaller than $c_{\rm{min}}$. The observed wave pattern
is in remarkable agreement with the theoretical prediction $r = a
\phi + b$ with $a$ given by Eq.~(\ref{rspiral}) and $r_0$ a free
parameter corresponding to an overall rotation of the spiral
\cite{Footnote}. For $V/c_{\rm{min}}$ lower than 0.8, no wake was
observed by naked eye. At lower rotation velocities, we probed the
surface deformation by measuring the deflexion of a laser beam
reflected by the air-water interface at a distance $r=11$~cm from
the rotation axis.

Using this scheme, we have established the existence of waves down
to $V/c_{\rm{min}} \approx 0.6$, and verified quantitatively that
the wave packet spectrum is peaked around $<\omega> \sim \kappa
\mathcal{R} \; \Omega$ (see Fig.~3). Experimentally, the frequency
$<\omega>$ corresponds to the period of the fast temporal
oscillations of the laser deflection angle (see Fig.~3 inset). In
order to compare our experimental results with our model, we  note
that the deflection of the laser at a point $\mathbf{r}$ is
proportional to the derivatives $\frac{1}{r}\frac{\partial
\zeta(\mathbf{r}, t)}{\partial \phi}$ and $\frac{\partial
\zeta(\mathbf{r}, t)}{\partial r}$. For simplicity, we will mainly
consider the angular derivative, but we have checked numerically
that our result do not depend on this choice. Using
Eq.~(\ref{zetaexpr}) the angular derivative can be decomposed into
Fourier series: $\frac{\partial \zeta(\mathbf{r}, t)}{\partial
\phi} = \sum_n C_n e^{i n (\phi
  - \Omega t) -i k_n r }$.
The coefficients $C_n$ are proportional to the contribution of the
frequency $n \Omega$ to the wave packet spectrum and we can thus
calculate the mean wave packet frequency using the expression:
$<\omega> = \Omega \sum_{n>0} n |C_n| / \sum_{n>0}  |C_n|$. As
shown in Fig.~3, our model is consistent with good accurcy with
the experimental data without any adjustable parameters.

Below $V/c_{\rm{min}} \approx 0.6$, the signal to noise ratio of
the experiment becomes to small to observe the laser deflection.
Note that this value is in qualitative agreement with Fig. 1 where
the wave resistance (hence the wave amplitude) has also
significatively decreased with respect to its maximum value for
$V/{c_{\rm{min}}}\lesssim 0.5$: we indeed note that for

\begin{figure}
\begin{center}
\includegraphics[width=0.9 \columnwidth]{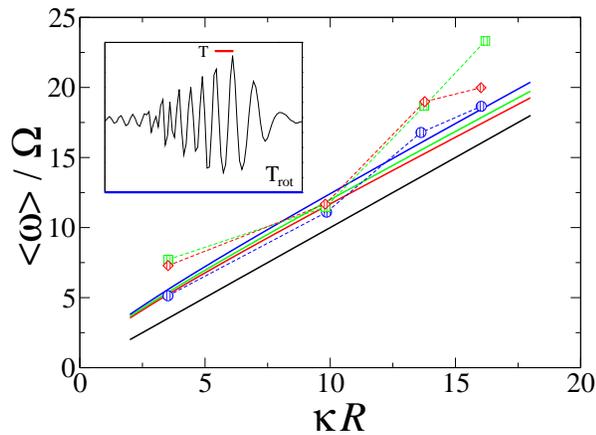}
\end{center}
\label{wavefig4} \caption{(Color online) Inset: typical time
dependence of the laser deflection angle (arbitrary units) during
a rotation period $T_{rot} = 2 \pi / \Omega$, the fast oscillation
frequency is given by $<\omega> = 2 \pi / T$. Main figure:
Dependence of the ratio $<\omega> / \Omega$ on $\kappa
\mathcal{R}$ for different needle velocities. The dashed curves
represent experimental results, while the continuous curve display
the numerical results of our model. Red, green and blue curves
(diamonds, squares and circles respectively) correspond to $V /
c_{\rm{min}} = 0.69, \; 0.76$ and $0.84$. The black curve
correspond to the analytical estimate $<\omega> / \Omega = \kappa
\mathcal{R}$. }
\end{figure}

To summarize, we have shown theoretically that a disturbance
moving along a circular trajectory experienced a wave drag even at
angular velocities corresponding to $V < c_{\rm{min}}$, where
$c_{\rm{min}}$ is the minimum phase velocity of capillary-gravity
waves. Our prediction is supported by experimental observation of
a long distance wake for $V/c_{\rm{min}}$ as low as $0.6$. For
$V/c_{\rm{min}} > 0.8$, we observed by naked eye Archimedean
spiral shaped crests, in good agreement with theory. These results
are directly related to the accelerated nature of the circular
motion, and thus do not contradict the commonly accepted threshold
$V = c_{\rm{min}}$ that is only valid for a rectilinear uniform
motion, an assumption often overlooked in the literature. It would
be very interesting to know if whirligig beetles can take
advantage of such spirals for echolocation purposes. Although
restricted to stationary wakes and thus excluded effects such as
vortex shedding, the results presented in this letter should be
important for a better understanding of the propulsion of
water-walking insects \cite{Alexander,Bush, Denis2,Buhler} where
accelerated motions frequently occurs ({\it{e.g}} when hunting a
prey or escaping a predator \cite{Bendele}). Even in the case
where the insect motion is rectilinear and uniform, one has to
keep in mind that the rapid leg strokes are accelerated and might
produce a wave drag even below $c_{\rm{min}}$.

We are grateful to Jos\'e Bico, J\'er\^ome Casas, M.~W. Denny and
J. Keller for fruitful discussions. F.C. acknowledges support from
R\'egion Ile de France (IFRAF) and A.C. acknowledges support from
Ecole Normale Sup\'erieure Paris.


\begin{thebibliography}{99}
\bibitem{FourierDefinition} The Fourier transform $\hat f(\boldsymbol{k},t)$
  is related with the function $f(\boldsymbol{r}, t)$ through
$f(\boldsymbol{r},t) = \int \frac{d^2 k}{(2 \pi)^2} e^{i \boldsymbol{k} . \mathbf{r}} \hat f(\boldsymbol{k},t)$
\bibitem{PressureField} While all the data shown in this letter are obtained with this
expression for $P_{ext}(r)$, we have verified that other
distributions (gaussian, step function, ...) lead qualitatively to
the same results.
\bibitem{Footnote} Note  that we also observed on some pictures the weaker spiral seen in Fig. 2 and propagating
in the opposite direction with respect to the main wake. However,
this ``advanced" wake could only be observed at high velocities
($V/c_{\rm{min}}\sim 2$) because of its small relative amplitude.
\end{thebibliography}
\end{document}